\documentclass[openacc]{rstransa}%%%%where rstrans is the template name

\usepackage{amsmath}
\usepackage{amsfonts}
\usepackage{amssymb}
\usepackage{bm}
\usepackage{mhchem}
\usepackage{subcaption}
\usepackage{sidecap}
\usepackage{cancel}

\usepackage[
backend=biber,
style=numeric,sorting=none,
doi=false,isbn=false,url=false,eprint=false]{biblatex}
\AtNextBibliography{\small}

\begin{document}

%%%% Article title to be placed here
\title{Reactive mixtures with the lattice Boltzmann model}

\author{N. Sawant, B. Dorschner and I. V. Karlin}

%%%%%%%%% Insert author address here
\address{Department of Mechanical and Process Engineering, ETH Zurich, 8092 Zurich, Switzerland}

%%%% Subject entries to be placed here %%%%
\subject{xxxxx, xxxxx, xxxx}

%%%% Keyword entries to be placed here %%%%
\keywords{xxxx, xxxx, xxxx}

%%%% Insert corresponding author and its email address}
\corres{I. V. Karlin \\
\email{ikarlin@ethz.ch}}

%%%% Abstract text to be placed here %%%%%%%%%%%%
\begin{abstract}
%200 word limit on the abstract. Currently 173. No citations in the abstract.
A new lattice Boltzmann model for reactive ideal gas mixtures is presented. 
The model is an extension to reactive flows of the recently proposed multi-component lattice Boltzmann model for compressible ideal gas mixtures with Stefan-Maxwell diffusion for species interaction. 
First, the kinetic model for the Stefan--Maxwell diffusion is enhanced to accommodate a source term accounting the change of the mixture composition due to chemical reaction. Second, by including the heat of formation in the energy equation, the thermodynamic consistency of the underlying compressible lattice Boltzmann model for momentum and energy allows a realization of the energy and temperature change due to chemical reactions. 
This obviates the need for ad-hoc modelling with source terms for temperature or heat. 
Both parts remain consistently coupled through mixture composition, momentum, pressure, energy and enthalpy. 
The proposed model uses the standard three-dimensional lattices and is validated with a set of benchmarks including laminar burning speed in the hydrogen-air mixture and circular expanding premixed flame.
\end{abstract}
%%%%%%%%%%%%%%%%%%%%%%%%%%%

%%%%%%%%%% Insert the texts which can accomdate on firstpage in the tag "fmtext" %%%%%

\begin{fmtext}
\section{Outline}
In this paper, we present derivation and analysis of the kinetic equations as well as the lattice Boltzmann formulation for Stefan--Maxwell diffusion for reactive mixtures. 
Subsequently, the compressible lattice Boltzmann model is extended to reactive flows. Finally, the model is validated for a set of benchmarks ranging from flame speed simulations of premixed hydrogen-air mixtures to challenging two-dimensional simulations of outward propagating circular flames with detailed chemistry.
\end{fmtext}
%%%%%%%%%%%%%%% End of first page %%%%%%%%%%%%%%%%%%%%%

\maketitle 

\section{Introduction}
The lattice Boltzmann method (LBM) is a recast of fluid dynamics into a fully discrete kinetic system for the populations $f_i(\bm{x},t)$ of designer particles, which are associated with the discrete velocities $\bm{c}_i$ fitting into a regular space-filling lattice. As a result, the kinetic equations for the populations $f_i(\bm{x},t)$ follow a simple algorithm of ``stream along links $\bm{c}_i$ and collide at the nodes $\bm{x}$ in discrete time $t$". 
{LBM has been successfully applied to a range of problems in fluid dynamics including but not limited to transitional flows, flows in complex moving geometries compressible flows, multiphase flows and rarefied gas, to name a few \cite{kruger_lattice_2017,succi_lattice_2018}.}

Nevertheless, in spite of extensive development, the multicomponent reactive mixtures so far resisted a significant advancement in the LBM context. 
Arguably, one of the main reasons was the absence of a thermodynamically consistent LBM for mixtures. Early approaches such as \cite{kang_lattice_2006,chiavazzo_combustion_2009} suffer many limitations such as  incompressible flow restriction, constant transport properties, rudimentary diffusion modelling.
As a remedy, a number of recent works  \cite{feng_lattice-boltzmann_2018,hosseini_hybrid_2019,tayyab_hybrid_2020} abandoned the construction of a kinetic model or LBM for multicomponent mixtures infavour of a so-called hybrid LBM where only the flow of the mixture is represented by an (augmented) LBM equation while the species and the temperature dynamics are modelled by conventional macroscopic equations. While the hybrid LBM approach can be potentially useful, in particular for combustion applications, our goal here is to retain a fully kinetic model and LBM for multicomponent reactive mixtures. 

Recently, we proposed a novel lattice Boltzmann framework for compressible multi-component mixtures with a realistic equation of state and thermodynamic consistency \cite{sawant_consistent_2021}. The strongly coupled formulation consists of kinetic equations for momentum, energy and species dynamics and 
was validated for a variety of test cases involving uphill diffusion, opposed jets and Kelvin-Helmholtz instability.
This extends the LBM to realistic mixtures and opens the door for reactive flow applications with a fully kinetic approach, which is the subject of this paper.
We propose a fully kinetic, strongly coupled lattice Boltzmann model for
compressible reactive flows as an extension of \cite{sawant_consistent_2021}.
To that end, a generic $M$-component ideal gas mixture is represented by two sets of kinetic equations. A set of $M$ kinetic equations is used to model species undergoing Stefan--Maxwell diffusion is  extended to include the reaction source term. 
Furthermore, the mixture is described by a set of two kinetic equations, where one accounts for the total mass and momentum of the mixture and another one for the total energy of the mixture. 
The kinetic equation for the mixture energy is extended to also include the internal energy of formation in addition to the sensible internal energy. 
Thus, the approach presented here can accurately model a reactive $M$-component compressible mixture with $M+2$ kinetic equations. The system is fully coupled through mixture composition, momentum, pressure, and enthalpy. The thermodynamic consistency of the model allows us to automatically account for the energy changes due to chemical reactions. The Stefan--Maxwell diffusion is retained and thus complicated phenomena such as reverse diffusion, osmotic diffusion or diffusion barrier can be captured, as it was already demonstrated in the non-reactive case in \cite{sawant_consistent_2021}.

The outline of the paper is as follows. In sec.\ \ref{sec:stefanMaxwell}, we extend the lattice Boltzmann model of Ref.\ \cite{sawant_consistent_2021} to the reactive multicomponent mixtures. This is achieved by supplying a reaction source term to the kinetic equations for the species in such a way that the Stefan--Maxwell diffusion mechanism already implemented by the model \ref{sec:stefanMaxwell} stays intact. In sec.\ \ref{sec:mix}, we extend the two-population lattice Boltzmann model for the mixture flow and energy to include the enthalpy of formation of chemically reacting species. Thanks to the thermodynamic consistency featured by the original model \cite{sawant_consistent_2021}, this final step completes the construction of the lattice Boltzmann model for the reactive mixtures. The derivation follows the path presented in detail in \cite{sawant_consistent_2021}, and we indicate the differences brought about by the thermodynamics of the chemical reaction. In sec.\ \ref{sec:cantera}, we outline the coupling of the lattice Boltzmann solver with the open source chemical kinetics package Cantera. Validation of the model is presented in sec.\ \ref{sec:results} with the simulation of detailed hydrogen/air combustion mechanism and the discussion is provided in sec.\ \ref{sec:conclusion}.

\section{Lattice Boltzmann model for the species}
\label{sec:stefanMaxwell}

The composition of a reactive mixture of $M$ ideal gases is described by the species densities $\rho_a$, $a=1,\dots, M$, while the mixture density is
$\rho=\sum_{a=1}^{M}\rho_a$.
The rate of change of $\rho_a$ due to chemical reaction $\dot \rho_a^c$ satisfies mass conservation,
 \begin{equation}
 \sum_{a=1}^{M}\dot \rho_a^c = 0.
 \label{eq:sumRhoDot}
 \end{equation}
Introducing the mass fraction
$ Y_a={\rho_a}/{\rho}$, the  molar mass of the mixture $m$ is given by
$
{m}^{-1}=\sum_{a=1}^M Y_a/m_a, 
$
where $m_a$ is the molar mass of the component $a$.
The equation of state of the mixture provides a relation between the pressure $P$, the temperature $T$ and the composition,
\begin{equation}
P=\rho R T,
\label{eq:eosIdealGas}
\end{equation}
where $R={R_U}/{m}$ is the specific gas constant of the mixture and $R_U$ is the universal gas constant. The pressure of an individual component $P_a$ is related to the pressure of the mixture $P$ through Dalton's law of partial pressures, 
$P_a=X_a P$, where the mole fraction of a component $X_a$ is related to its mass fraction $Y_a$ as $X_a={m} Y_a /{m_a}$.
Combined with the equation of state (\ref{eq:eosIdealGas}), the partial pressure $P_a$ takes the form $P_a=\rho_a R_a T$,
where $R_a={R_U}/{m_a}$ is the specific gas constant of the component.

Kinetic model for the Stefan--Maxwell diffusion in the non-reactive mixture were introduced in \cite{sawant_consistent_2021}. Here, we extend the formulation \cite{sawant_consistent_2021} to include the reaction.
To that end, we write the kinetic equation for the populations $f_{ai}$, $a=1,\dots,M$, of the component $a$, corresponding to the discrete velocities $\bm{c}_i$, $i=0,\dots, Q-1$,
\begin{equation}
\partial_t f_{ai} + \bm{c}_{i}\cdot \nabla f_{ai} = \sum_{b\ne a}^M \frac{PX_aX_b}{\mathcal{D}_{ab}} \left[ \left( \frac{f_{ai}^{\rm eq}-f_{ai}}{\rho_a} \right) - \left( \frac{f_{bi}^{\rm eq}-f^*_{bi}}{\rho_b} \right) \right]+ \dot f_{ai}^c.
\label{eq:stefanMaxwell} 
\end{equation}
Here $\mathcal{D}_{ab}$ are the binary diffusivity coefficients. The species' densities $\rho_a$ and partial momenta $\rho_a \bm{u}_a$ are, respectively,
\begin{align}
\rho_a= \sum_{i=0}^{Q-1}f_{ai},
\quad
\rho_a \bm{u}_a= \sum_{i=0}^{Q-1} f_{ai}\bm{c}_i.
\label{eq:Pa}
\end{align}
The momenta of the components sum up to the mixture momentum,
At variance with the non-reactive mixture \cite{sawant_consistent_2021}, kinetic equation (\ref{eq:stefanMaxwell}) includes a source term $\dot f_{ai}^c$ which implements the rate of change of $\rho_a$ due to the reaction and satisfies the following conditions,
\begin{align}
\sum_{i=0}^{Q-1} \dot f_{ai}^{c} =  \dot \rho_a^c, \quad
\sum_{i=0}^{Q-1} \dot f_{ai}^{c}\bm{c}_i &=  \dot \rho_a^c \bm{u}.
\label{eq:sumfDotc}
\end{align}
The kinetic model (\ref{eq:stefanMaxwell}) is realized on the standard three-dimensional $D3Q27$ lattice with the discrete velocities $\bm{c}_i=(c_{ix},c_{iy},c_{iz}),\ c_{i\alpha}\in\{-1,0,1\}$.
Same as in \cite{sawant_consistent_2021}, the equilibrium $f_{ai}^{\rm eq}$ and the quasi-equilibrium $f_{ai}^*$ in (\ref{eq:stefanMaxwell}) are constructed using the product-form \cite{karlin_factorization_2010}: We define a triplet of functions in two variables, $\xi$ and $\zeta>0$,
\begin{align}
	\Psi_{0}(\xi,\zeta) = 1 - (\xi^2 + \zeta), \
		\label{eqn:phis}
	%	\\
	\Psi_{1}(\xi,\zeta) = \frac{\xi + (\xi^2 + \zeta)}{2},\ 
	%	\label{eqn:phiPlus}
	%	\\
	\Psi_{-1}(\xi,\zeta) = \frac{-\xi + (\xi^2 + \zeta)}{2}.
	%	\label{eqn:phiMinus}
\end{align}
The equilibrium $f_{ai}^{\rm eq}$ and the quasi-equilibrium $f_{ai}^{*}$ populations are evaluated as the products of the functions (\ref{eqn:phis}), with $\xi=u_{\alpha}$ and {$\xi=u_{a\alpha}$}, respectively, and with $\zeta=R_aT$ in both cases,
\begin{align}
	&f_{ai}^{\rm eq}(\rho_a,\bm{u},R_aT)=  \rho_a  \Psi_{c_{ix}}\left(u_{x},R_aT\right)
	\Psi_{c_{iy}}\left(u_{y},R_aT\right) 
	\Psi_{c_{iz}}\left(u_{z},R_aT\right),
	\label{eq:feq} \\
	&f_{ai}^{*}(\rho_a,\bm{u_a},R_aT)= 
\rho_a  \Psi_{c_{ix}}\left(u_{ax},R_aT\right)
	\Psi_{c_{iy}}\left(u_{ay},R_aT\right) 
	\Psi_{c_{iz}}\left(u_{az},R_aT\right). 
\label{eq:fstar} 	
\end{align}
The reaction source term $\dot f_{ai}^c$ in (\ref{eq:stefanMaxwell}) is also represented with the product-form similar to (\ref{eq:feq}),
\begin{align}
	&\dot f_{ai}^{c}(\dot \rho_a^c,\bm{u},R_aT)= \dot \rho_a^c  \Psi_{c_{ix}}\left(u_{x},R_aT\right)
\Psi_{c_{iy}}\left(u_{y},R_aT\right) 
\Psi_{c_{iz}}\left(u_{z},R_aT\right).
\label{eq:fsource}
\end{align}
The analysis of the hydrodynamic limit of the kinetic model (\ref{eq:stefanMaxwell}) follows the lines already presented in \cite{sawant_consistent_2021}. Note that the constraint on the momentum of the source term (\ref{eq:sumfDotc}) is required. The balance equations for the densities of the species in the presence of the source term are found as follows,
\begin{align}
	\partial_t\rho_a&=-\nabla\cdot(\rho_a \bm{u})-\nabla\cdot(\rho_a \delta\bm{u}_a) + \dot \rho_a^c,
	\label{eq:dtrhoa}
\end{align}
where the diffusion velocities, $\delta\bm{u}_a=\bm{u}_a-\bm{u}$, satisfy the Stefan--Maxwell constitutive relation,
\begin{equation}
	P\nabla X_a+(X_a-Y_a)\nabla P=\sum_{b\ne a}^M \frac{PX_aX_b}{\mathcal{D}_{ab}} \left({\delta}\bm{u}_{b} - {\delta}\bm{u}_a \right).
	\label{eq:constit2}
\end{equation}
Summarizing, kinetic model (\ref{eq:stefanMaxwell}) recovers both the Stefan--Maxwell law of diffusion
	%, with the diffusion due to non-uniformity of the species molar concentration and barodiffusion
	 and the contribution of the species mass change due to chemical reaction, as presented in equation (\ref{eq:dtrhoa}).

Derivation of the lattice Boltzmann equation from the kinetic model (\ref{eq:stefanMaxwell}) proceeds along the lines of the non-reactive case \cite{sawant_consistent_2021}. Upon integration of \eqref{eq:stefanMaxwell} along the characteristics and application of the trapezoidal rule, we arrive
at a fully discrete lattice Boltzmann equation,
\begin{equation}
 f_{ai}(\bm{x}+\bm{c}_i \delta t, t+ \delta t)  = f_{ai}(\bm{x},t)+ 2 \beta_a [f_{ai}^{\rm eq}(\bm{x},t) - f_{ai}(\bm{x},t)]
+ \delta t (\beta_a-1) F_{ai}(\bm{x}, t) + \delta t \dot f_{ai}^c.
\label{eq:finalNumericalEquations}
\end{equation}
The shorthand notation $F_{ai}$ for the inter-species interaction term and the relaxation parameters $\beta_a\in[0,1]$ are,
\begin{equation}
F_{ai} = Y_a \sum_{b\ne a}^M \frac{1}{\tau_{ab}}  \left( f_{bi}^{\rm eq}-f_{bi}^* \right), \quad
\beta_a=\frac{\delta t}{2 \tau_a + \delta t}, 
\label{eqn:fStar}
\end{equation}
where the characteristic times $\tau_{ab}$ and the relaxation times $\tau_a$ are related to the binary diffusivities,
\begin{equation}
	{\tau_{ab}} = \left(\frac{m_a m_b}{mR_UT}\right)\mathcal{D}_{ab}, \quad
	%\left(\frac{m}{m_a m_b}\right), \ 
	\frac{1}{\tau_a} = \sum_{b\ne a}^M \frac{Y_b}{\tau_{ab}}.
	% =R_a T\left(\sum_{b\ne a}^M \frac{X_b}{\mathcal{D}_{ab}}\right).
	\label{eqn:tauab}
\end{equation}
Furthermore, the quasi-equilibrium populations $f_{bi}^*=f_{bi}^{*}( \rho_b,\bm{u}+\delta \bm{u}_b,R_bT)$ in the expression $F_{ai}$ (\ref{eqn:fStar}) depend on the diffusion velocity $\delta \bm{u}_b$. The latter are found by solving the $M\times M$ linear algebraic system for each spatial component,
\begin{align}
\left( 1+ \frac{\delta t}{2 \tau_a}\right) \delta \bm{u}_{a} - \frac{\delta t}{2} \sum_{b\ne a}^{M} \frac{1}{\tau_{ab}} Y_b \delta \bm{u}_{b}=\bm{u}_{a}-\bm{u}.
\label{eqn:transform1}
\end{align}
The linear algebraic system was already derived in \cite{sawant_consistent_2021} for the non-reactive mixtures and is not altered by the presence of the reaction source term. The equilibrium population $f_{ai}^{eq}=f_{ai}^{eq}( \rho_a,\bm{u},R_aT)$ and the reaction source term $\dot f_{ai}^c=\dot f_{ai}^{c}(\dot \rho_a,\bm{u},R_aT)$ in (\ref{eq:finalNumericalEquations}) and (\ref{eqn:fStar}) are evaluated at the mixture velocity $\bm{u}$.
Summarizing, the lattice Boltzmann system (\ref{eq:finalNumericalEquations}) delivers the extension of the species dynamics subject to the Stefan--Maxwell diffusion to the reactive mixtures. We proceed with the extension of the flow and energy dynamics of the mixture.

\section{Lattice Boltzmann model of mixture momentum and energy}
\label{sec:mix}
The mass-based specific internal energy ${U}_{a}$ and enthalpy ${H}_{a}$ of a specie $a$ are,
\begin{align}
{U}_{a}=U^0_a+\int_{T_0}^T{C}_{a,v}(T')dT',
 \quad
{H}_{a}=H^0_a+\int_{T_0}^T{C}_{a,p}(T')dT',
\label{eq:specHa}
\end{align}
where $U^0_a$ and  $H^0_a$ are, respectively, the energy and the enthalpy of formation at the reference temperature $T_0$, while $C_{a,v}$ and $C_{a,p}$ are specific heats at constant volume and at constant pressure.
The internal energy $\rho U$ and the enthalpy $\rho H$ of a mixture are,
\begin{align}
\rho U=\sum_{a=1}^M\rho_a U_a,\quad \rho H=\sum_{a=1}^M\rho_a H_a.\label{eq:Umix}
\end{align}
While the sensible heat was considered in the non-reactive case \cite{sawant_consistent_2021}, by taking into account the heat of formation we immediately extend the model to reactive mixtures.	
Same as in \cite{sawant_consistent_2021}, we follow a two-population approach.
One set of populations ($f$-populations) is used to represent the density and the momentum of the mixture, 
\begin{align}
\sum_{i=0}^{Q-1} f_i  = \rho,\quad \sum_{i=0}^{Q-1} f_i \bm{c}_{i} = \rho \bm{u}.
\end{align}
Another set ($g$-populations)  represents the total energy,
\begin{align}
\sum_{i=0}^{Q-1} g_i  =  \rho E,\quad \rho E=\rho U + {\frac{\rho u^2}{2}}. \label{eq:g0momTotalEnergy}
\end{align}
%%
%\begin{align}
%	\rho E=\rho U + {\frac{\rho u^2}{2}}.
%\end{align}
A coupling between the mixture and the species kinetic equations is established through energy since the mixture internal energy (\ref{eq:Umix}) depend on the composition. 
Furthermore, the temperature is evaluated by solving the integral equation, cf.\ (\ref{eq:specHa}) and (\ref{eq:Umix}),
\begin{equation}
	\label{eq:temperature}
	\sum_{a=1}^MY_a\left[U_a^0 + \int_{T_0}^T {C}_{a,v}(T')dT'\right]=E-\frac{u^2}{2}.
\end{equation}
The temperature is used as the input for the equation of state (\ref{eq:eosIdealGas}) and hence in the equilibrium, the quasi-equilibrium and the reaction source term of the species lattice Boltzmann system which leads to a two-way coupling between the species and the mixture kinetic systems.  
Same as in \cite{sawant_consistent_2021}, the lattice Boltzmann equations for the $f$- and $g$-populations are realized on the $D3Q27$ discrete velocity set, 
\begin{align}
f_i(\bm{x}+\bm{c}_i \delta t,t+\delta t)- f_i(\bm{x},t)&=  \omega (f_i^{\rm eq} -f_i)+ \bm{A}_{i}\cdot \bm{X},
\label{eq:f} 
\\
g_i(\bm{x}+\bm{c}_i \delta t,t+ \delta t) - g_i(\bm{x},t)&=  \omega_1 (g_i^{\rm eq} -g_i) + (\omega - \omega_1) (g_i^{*} -g_i),
 \label{eq:g}
\end{align}
where relaxation parameters $\omega$ and $\omega_1$ are related to the viscosity and thermal conductivity. 
The equilibrium $f$-populations $f_i^{\rm eq}$ in (\ref{eq:f}) are evaluated using the product-form, with ${\xi}_{\alpha}={u}_{\alpha}$ and $\zeta=RT$ in (\ref{eqn:phis}),
\begin{equation}
f_{ai}^{\rm eq}(\rho,\bm{u},RT)=  \rho \Psi_{c_{ix}}\left(u_{x},RT\right)
\Psi_{c_{iy}}\left(u_{y},RT\right) 
\Psi_{c_{iz}}\left(u_{z},RT\right).
\label{eq:feqmix}
\end{equation}
The last term in (\ref{eq:f}) is a correction needed to compensate for the insufficient isotropy of the $D3Q27$ lattice in the compressible flow setting \cite{saadat_lattice_2019,sawant_consistent_2021}: $\bm{X}$ is the vector with the components,
\begin{equation}
	X_{\alpha} = -\partial_\alpha \left[ \left( \frac{1}{\omega}-\frac{1}{2} \right)\delta t
	\partial_\alpha (\rho u_\alpha (1 - 3 R T) - \rho u_\alpha^3) \right],
	\label{eqn:Xalpha}
\end{equation}
while the components of vectors $\bm{A}_i$ are defined as,
\begin{equation} \label{eqn:Ai}
		A_{i \alpha}  = \frac{1}{2} c_{i \alpha}\ \  \text{for} \;  c_i^2=1;\quad  
		A_{i \alpha}  = 0\ \ \text{otherwise}.  
\end{equation}
The equilibrium and the quasi-equilibrium $g$-populations, $g_i^{\rm eq}$ and $g_i^{*}$ in (\ref{eq:g}), are defined with the help of Grad's approximation \cite{grad_kinetic_1949},
	\begin{align}
		g_i^{\rm eq}&= w_i \left(\rho E +  \frac{ \bm{q}^{\rm eq}\cdot \bm{c}_{i}}{\theta} + \frac{(\bm{R}^{\rm eq}-\rho E \theta \bm{I}): (\bm{c}_{i}\otimes\bm{c}_{i} - \theta \bm{I})}{2 \theta^2}\right),
		\label{eq:gieq}\\
		g_i^{*}&= w_i \left(\rho E +  \frac{ \bm{q}^{*}\cdot \bm{c}_{i}}{\theta} + \frac{(\bm{R}^{\rm eq}-\rho E \theta \bm{I}): (\bm{c}_{i}\otimes\bm{c}_{i} - \theta \bm{I})}{2 \theta^2}\right),
		\label{eq:giqeq}
	\end{align}
Here, the weights $w_i = w_{c_{ix}} w_{c_{iy}} w_{c_{iz}}$ are the products of the one-dimensional weights $w_{0} = 1 - \theta$, $w_{1} = w_{-1} = {\theta}/{2}$, and  $\theta=1/3$ is the lattice reference temperature.
The equilibrium mixture energy flux  $\bm{q}^{\rm eq}$ and the second-order moment tensor $\bm{R}^{\rm eq}$ in (\ref{eq:gieq}) and (\ref{eq:giqeq}) are,
\begin{align}
\bm{q}^{\rm eq}&= \sum_{i=0}^{Q-1}  g_i^{\rm eq} \bm{c}_{i} =  \left(H+\frac{u^2}{2}\right)\rho\bm{u},
\label{eq:geq1mom}
\\
\bm{R}^{\rm eq}&=\sum_{i=0}^{Q-1} g_i^{\rm eq} \bm{c}_i\otimes\bm{c}_i =
    \left(H+\frac{u^2}{2}\right) \bm{P}^{\rm eq} + P\bm{u}\otimes\bm{u},
\label{eq:geq2mom}
\end{align}
%
%The zeroth and the second-order moments of the quasi-equilibrium populations $g_i^{*}$ are the same as that of the equilibrium population $g_i^{eq}$. 
where $H$ is the specific mixture enthalpy (\ref{eq:Umix}). The quasi-equilibrium energy flux $\bm{q}^*$ in (\ref{eq:giqeq}) has the following form,
\begin{align}
\bm{q}^{*} =\sum_{i=0}^{Q-1} g_i^{*} \bm{c}_{i} =  \bm{q} -  \bm{u}\cdot (\bm{P} - \bm{P}^{\rm eq}) +\bm{q}^{\rm diff}+\bm{q}^{\rm corr}.
\label{eq:gstareq1mom}
\end{align}
The two first terms in (\ref{eq:gstareq1mom}) include the energy flux $\bm{q}$ and the pressure tensor $\bm{P}$,
\begin{align}
	\bm{q}=\sum_{i=0}^{Q-1} g_i \bm{c}_{i},\quad \bm{P}=\sum_{i=0}^{Q-1} f_i \bm{c}_{i}\otimes \bm{c}_{i}.
\end{align}
Their contribution maintains a variable Prandtl number and is patterned from the single-component case \cite{saadat_lattice_2019}.
The remaining two terms in the quasi-equilibrium energy flux (\ref{eq:gstareq1mom}), $\bm{q}^{\rm diff}$ and $\bm{q}^{\rm corr}$ pertain to the multicomponent case. The interdiffusion energy flux $\bm{q}^{\rm diff}$ is,
\begin{align}
\bm{q}^{\rm diff} =\left(\frac{\omega_1}{\omega-\omega_1} \right) \rho\sum_{a=1}^{M}H_aY_a \delta\bm{u}_a,\label{eq:qdiff}
\end{align}
where the diffusion velocities $\delta\bm{u}_a$ are defined according to Eq.\ (\ref{eqn:transform1}).
%(\ref{eq:perturb}). 
The flux (\ref{eq:qdiff}) contributes the enthalpy transport due to diffusion and hence it vanishes in the single-component case but is significant in reactive flows.
Finally, the correction flux $\bm{q}^{\rm corr}$, which also vanishes in the single-component case, is required in the two-population approach to the mixtures in order to recover the Fourier law of thermal conduction, see \cite{sawant_consistent_2021} for details,
\begin{align}
\bm{q}^{\rm corr}=\frac{1}{2}\left(\frac{\omega_1-2}{\omega_1-\omega}\right) {\delta t}P \sum_{a=1}^M  H_{a} \nabla  Y_a.
\label{eq:corrFourier}
\end{align}
Prefactors featured in (\ref{eq:qdiff}) and (\ref{eq:corrFourier}) were found in \cite{sawant_consistent_2021} based on the analysis of the hydrodynamic limit of the lattice Boltzmann system (\ref{eq:f}) and (\ref{eq:g}) and are not affected by the present reactive mixture case. {Second-order accurate isotropic lattice operators proposed in \cite{thampi_isotropic_2013} were used for the evaluation of spatial derivatives in the correction flux (\ref{eq:corrFourier}) as well as in the isotropy correction (\ref{eqn:Xalpha}).} Following \cite{sawant_consistent_2021},  the continuity, the momentum and the energy equations for a reactive multicomponent mixture \cite{williams_combustion_1985} are obtained as follows, 
%
%\noindent The continuity equation:
\begin{align}
&\partial_t \rho + \nabla\cdot (\rho \bm{u})=0,
\label{eq:dtrho}\\
&\partial_t (\rho\bm{u}) +  \nabla\cdot ({\rho\bm{u}\otimes\bm{u} })+ \nabla\cdot \bm{\pi}=0,
\label{eq:dtu}\\
&\partial_t (\rho E)+\nabla\cdot(\rho E\bm{u})+\nabla\cdot\bm{q}+\nabla\cdot(\bm{\pi}\cdot\bm{u})=0.
\label{eq:dtE}
\end{align}
The pressure tensor $\bm{\pi}$ in the momentum equation (\ref{eq:dtu}) reads,
\begin{equation}\label{eq:NSmix}
\bm{\pi}=P\bm{I}
-\mu \left( \nabla\bm{u}  + \nabla\bm{u}^{\dagger}  -\frac{2}{D} (\nabla\cdot\bm{u})\bm{I} \right) 
-\varsigma (\nabla\cdot\bm{u}) \bm{I}, 
\end{equation}
%
%%%%%%%%%%%%%%%%%%% inset viscosity 
where the dynamic viscosity $\mu$ and the bulk viscosity $\varsigma$ are related to the relaxation parameter $\omega$,
\begin{align}
	\mu  = \left( \frac{1}{\omega} - \frac{1}{2}\right) P{\delta t},\quad
	\label{eq:mu}
	\varsigma =\left( \frac{1}{\omega}-\frac{1}{2}\right)\left( \frac{2}{D} - \frac{R}{C_v} \right) P{\delta t},
	%\label{eqn:varsigma}
\end{align}
where $C_v=\sum_{a=1}^MY_aC_{a,v}$ is the mixture specific heat at constant volume.
%%%%%%%%%%%%%%%%%%%%%%%%%%%%%%%%%%%%%%%%%
The heat flux $\bm{q}$ in the energy equation (\ref{eq:dtE}) reads,
\begin{equation}
\label{eq:qneq}
\bm{q}=-\lambda\nabla T+\rho\sum_{a=1}^{M}H_aY_a \delta\bm{u}_a.
\end{equation}
The first term is the Fourier law of thermal conduction, with the thermal conductivity $\lambda$ related to the relaxation parameter $\omega_1$,
\begin{equation}\label{eq:lambda}
	\lambda= \left(\frac{1}{\omega_1} - \frac{1}{2}\right) P C_p{\delta t},
\end{equation}
where $C_p=C_v+R$ is the mixture specific heat at constant pressure. The second term in (\ref{eq:qneq}) is the interdiffusion energy flux.
The dynamic viscosity $\mu$ and the thermal conductivity $\lambda$ of the mixture are evaluated as a function of the local composition, temperature and pressure using the chemical kinetics solver Cantera \cite{goodwin_cantera_2018}, wherein a combination of methods involving interaction potential energy functions \cite{kee_chemically_2003}, hard sphere approximations and the methods described in \cite{wilke_viscosity_1950} and \cite{mathur_thermal_1967} are employed to calculate the mixture transport coefficients.
{Finally, in accord with a principle of strong coupling \cite{sawant_consistent_2021}, the excess conservation laws arising due to a separated construction of the species diffusion model in sec.\ \ref{sec:stefanMaxwell} and the two-population mixture model are eliminated by removing one set of species populations (here, the component $M$),
\begin{equation}\label{eq:strongest}
f_{Mi}=f_{i}-\sum_{a=1}^{M-1}f_{ai}.
\end{equation}
Thus, the component $M$ is not an independent field any more but is slaved to the remaining $M-1$ species and the mixture $f$-populations.} 
Summarizing, the thermodynamically consistent framework of \cite{sawant_consistent_2021} allows for a straightforward extension to reactive mixtures provided the sensible energy and enthalpy are extended to include the energy and the enthalpy of formation.

%\subsection{Realization on the standard lattice}
%\label{sec:realization}
%
\section{Coupling between lattice Boltzmann and chemical kinetics}
\label{sec:cantera}
In this work, the lattice Boltzmann code is coupled to the open source code chemical kinetics solver Cantera \cite{goodwin_cantera_2018}. The Cantera solver is supplied with the publicly accessible GRI-Mech 3.0 mechanism \cite{smith_gri-mech_1999} as an input data file. 
%A sub-mechanism can then be extracted from GRI mechanism involving only the species and reactions which constitute the 9 species mechanism of \cite{li_updated_2004}. 
The communication between the lattice Boltzmann solver and the Cantera chemical kinetics solver is executed as follows: 
\begin{enumerate}
    \item An input from the lattice Boltzmann solver to Cantera is provided during the collision step in terms of internal energy, specific volume and mass fractions. 
    \item Cantera internally solves numerically the integral equation (\ref{eq:temperature}) and thus the temperature at that state is obtained. 
    \item Cantera calculates the production rates of species $\dot \rho_a^c$ and the transport coefficients including dynamic viscosity, thermal conductivity and the Stefan--Maxwell diffusivities as a function of the current state.
    \item  The temperature obtained from Cantera is used to evaluate the equilibrium and quasi-equilibrium moments and populations. The transport coefficients are used to calculate the corresponding relaxation times and thus the collision step is complete.  
\end{enumerate}
Other thermodynamic parameters necessary for the simulations such as the specific heats and molecular masses are also obtained through Cantera. The reference standard state temperature is $T_0=298.15 K$ and the reference standard state pressure is $P_0=1\ atm$. The data required by the lattice Boltzmann solver during runtime is obtained by querying Cantera through its C++ API using the "IdealGasMix" and "Transport" classes.
\section{Results}
\label{sec:results}
As a first validation, probing the basic validity of our model,  we 
compute the flame speed in a premixed hydrogen/air mixture with the reactive Stefan--Maxwell formulation in a wide range of equivalence ratios $\phi$. 
Subsequently, in order to test the isotropy of the model, the problem of outward expanding circular flame \cite{altantzis_numerical_2013,altantzis_direct_2015} is solved for the premixed hydrogen/air mixture. For both test cases, we use the detailed chemical kinetics mechanism \cite{li_updated_2004} involving the following nine species: \ce{ N2, O2, H2, H, O, OH, H2O, HO2, H2O2}. It is worthwhile to mention that the model is not restricted to the detailed mechanisms. Reduced mechanisms available in the literature such as the five-species propane mechanism has also been tested with this model. In this paper, we will restrict ourselves to the more interesting detailed hydrogen/air mechanism which forms sharper and faster propagating flames. 
While this benchmark not only probes the model's behaviour in two dimensions, it is also a stringent isotropy test where it is crucial that the circular shape of the flame is preserved and not contaminated or distorted by the errors of the discrete numerics on the underlying Cartesian grid.
Finally, the models ability to capture non-linear instabilities is probed by simulations of wrinkled flames, which form as a result of 
polychromatic perturbations.

\subsection{Laminar flame speed}
\begin{figure}
	\centering
		\includegraphics[width=0.6\linewidth]{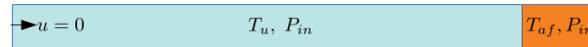}
		\caption{Setup for the $D=1$ burning velocity simulation.}
		\label{fig:flameSetup}
\end{figure}
\begin{figure}
	\centering
		\includegraphics[width=0.75\linewidth]{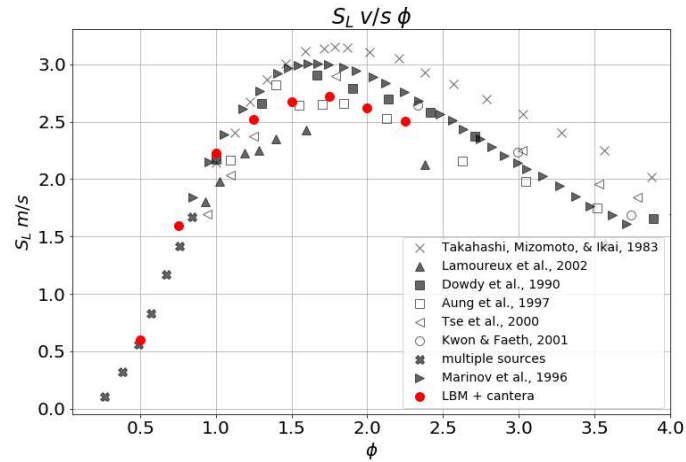}
		\caption{Burning velocity $S_L$ vs. equivalence ratio $\phi$ for the nine-species hydrogen/air mixture detailed chemistry. Reference: \cite{dahoe_laminar_2005} }
		\label{fig:flameSpeed}
\end{figure}

\begin{figure}[h]
\begin{subfigure}{0.5\textwidth}
\includegraphics[width=0.9\linewidth]{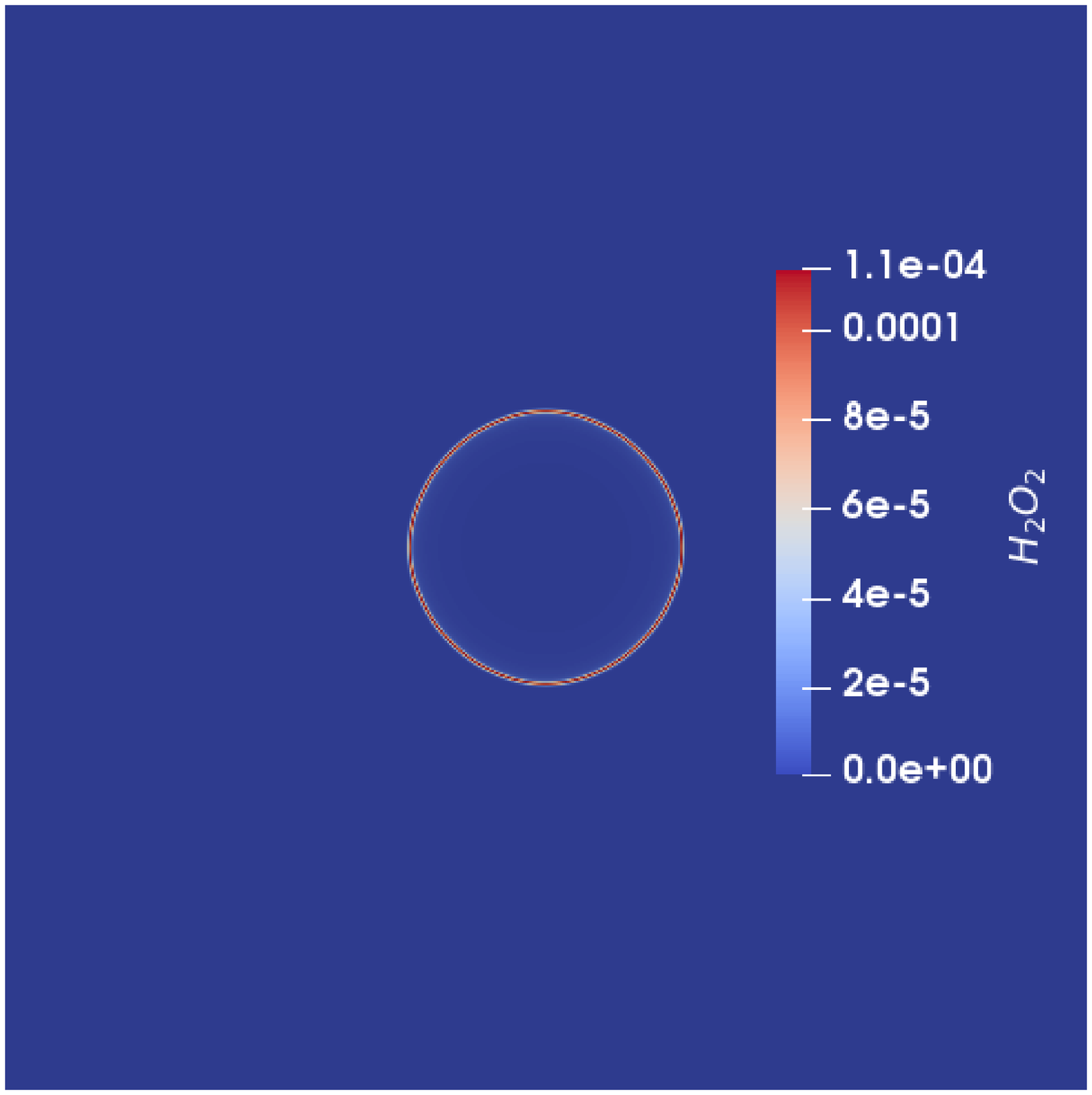} 
\caption{Contour plot of the mole fraction of $H_2O_2$ at t=$0.082 \tau$ obtained by reflecting about the left edge and the bottom edge of the domain.}
\label{fig:reflectedH2O2}
\end{subfigure}
\begin{subfigure}{0.5\textwidth}
\includegraphics[width=0.9\linewidth]{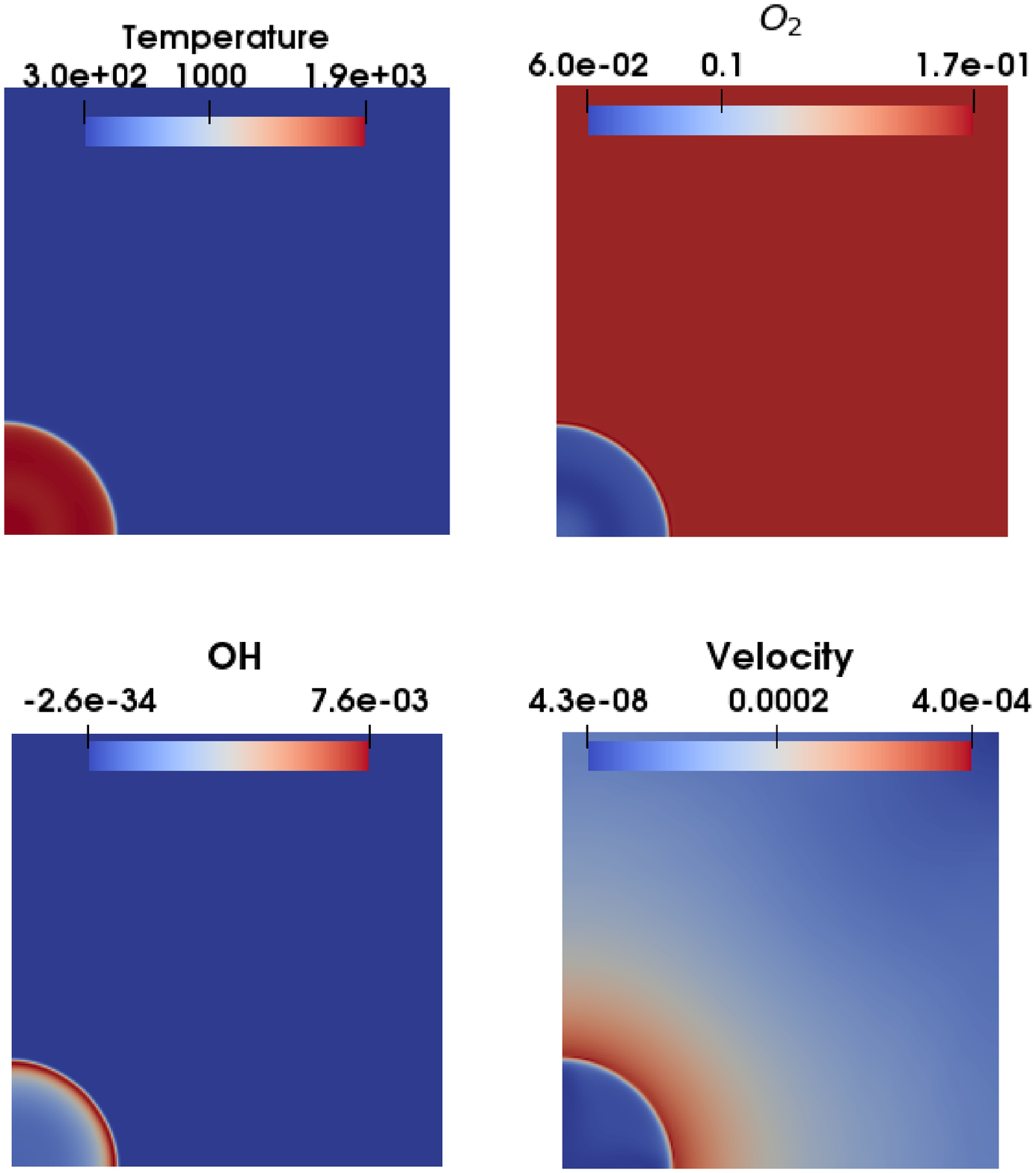}
\caption{Contours of temperature, mole fractions of $O_2$, $OH$ and velocity at $t=0.082 \tau$.}
\label{fig:circFlame}
\end{subfigure}
\caption{Premixed hydrogen/air circular outward expanding flame.}
\label{fig:image2}
\end{figure}
In order to validate our model, we calculate the burning velocity of a hydrogen/air mixture in a one-dimensional setup. As illustrated in Fig.\ \ref{fig:flameSetup}, the setup 
%is similar to \cite{chiavazzo_combustion_2009},
consists of a one-dimensional tube initialized with unburnt mixture at $T_{\rm u}=300 K$ throughout from the left end up to 80\% of the domain towards the right. The remaining 20\% of the domain are initialized with the adiabatic flame temperature $T_{\rm af}$ and with the equilibrium burnt composition at the respective equivalence ratio. The pressure is initialized uniformly at $P_{in}=1\, atm$. Zero gradient boundary conditions are used at both ends for all variables using equilibrium populations. 
%This is done by replacing before advection the populations at the pseudo nodes with equilibrium distribution calculated at the moments of the adjacent boundary populations. 
At the left end, the velocity is imposed to be zero so that the flame propagates from right to left into the stationary unburnt mixture. The setup is used to calculate the burning velocity of the premixed \ce{H2, N2, O2} system. Nitrogen is considered as an inert gas and thus does not split or form any radicals like nitrous oxides. However, the heat capacity of the inert gas has a strong influence on the flame temperature and consequently on the burning velocity. This is naturally accounted for in the formulation. The burning velocity is measured for various equivalence ratios ranging from $\phi=0.5$ to $\phi=2.25$. 
%
%The equivalence ratio is defined as the ratio of the fuel to air ratio to the stoichiometric fuel to air ratio:
%%
%\begin{align}
%    \phi = \frac{X_{H_2}/(X_{O_2}+X_{N_2})}{(X_{H_2}/(X_{O_2}+X_{N_2}))_{stoichiometric}}.
%\label{eq:defEqRatio}
%\end{align}
%%
%The global reaction for $\phi=1$ can be written as
%\begin{align}
%\ce{H_2 + \frac{1}{2} (O_2 + 3.67 N_2) -> H_2O + 1.88 N_2}.
%\label{eq:h2o2n2GlobalReaction}
%\end{align}
%%
%Thus, $\phi>1$ represents a fuel rich mixture while $\phi<1$ means a fuel deficient or a lean mixture. 
%%
We use the laminar flame thickness $\delta_{\rm f}$ at $\phi=1$ for defining the reference length, 
where $\delta_{\rm f}=(T_{\rm af}-T_{\rm u})/max(\left|{dT}/{dx}\right|)$.
In order to accurately calculate the burning velocity, we use a long domain of $N \approx 90 \delta_{\rm f}$, which corresponds to $10^4$ lattice points. In order to avoid the effect of the boundaries and transients due to initial acceleration, the flame speed $S_L$ is measured when the flame front approaches the middle of the domain. The results are compared to the data provided by \cite{dahoe_laminar_2005} from multiple experimental and computational sources in Fig.\ \ref{fig:flameSpeed}. It can be seen that flame speed computed by our model agrees well with the data available in the literature. Although there is considerable dispersion in the literature for the flame speed values for fuel-rich mixtures $\phi>1$, the location of the peak burning velocity between $\phi=1.5$ and $\phi=2.0$ has been correctly captured. This test case indicates that the present model is a promising candidate for simulating reactive flows with the lattice Boltzmann method.

\subsection{Circular expanding premixed flame}
After confirming the $1D$ behaviour of the model, we compute the $2D$ circular expanding flame in a premixed hydrogen/air mixture with detailed chemistry. Similarly to the study \cite{altantzis_numerical_2013,altantzis_direct_2015}, due to symmetry, only a quarter of the flame is solved. Symmetry boundary conditions are used on the left and bottom edges of the square domain while the characteristic based outlet boundary conditions \cite{poinsot_boundary_1992,feuchter_turbulent_2019} are imposed at the right and top edges of the domain. The bottom left corner is initialized with a burnt quarter sector at the adiabatic flame temperature $T_{\rm af}=1844.27 K$ corresponding to the equivalence ratio $\phi=0.6$. The rest of the domain is initialized with an unburnt mixture at the temperature $T_{\rm u}=298 K$. The composition in the burnt section is set to the equilibrium composition and the pressure in the entire domain is initialized to a uniform pressure  $P=5\, atm$. 
For this premixed initial condition, the burning velocity is obtained as $S_L=38.11 \text{ cm/s}$ from solving a $1D$ flame propagation setup in Cantera. The flame thickness at these initial conditions is obtained as $\delta_{\rm f}=8.8 \times 10^{-3}$ cm. A square domain with the side $N \approx 51 \delta_f$ was considered, which corresponds to $1200 \times 1200$ lattice points. The radius of the region initialized with the burnt equilibrium conditions is $R_{\rm ig} \approx 8.5 \delta_{\rm f}$. 
%A tangent hyperbolic profile is used at the initial interface of the burnt and the unburnt mixture to avoid numerical instability because of the staggered representation of the circular interface on a Cartesian grid.
 
The characteristic flame transit time is defined as $\tau={\delta_f}/{S_L}=2.31 \times 10^{-4}\, s$  \cite{altantzis_direct_2015}. Contours of temperature, velocity and mole fractions of oxygen and the hydroxide radical are shown at $t = 0.082 \tau$ in Fig.\ \ref{fig:circFlame}. As can be verified from Fig.\ \ref{fig:circFlame}, the solution is not contaminated by numerical noise or anisotropies and the contours do not contain any other spurious features. The thin interface of the hydroxide radical at the flame front is captured correctly and the curvature of the flame is maintained. This is in contrast to, e.g., \cite{altantzis_direct_2015}, where the errors of the underlying numerical discretization leading to a spurious behaviour were reported when using Cartesian grids.
  
\begin{figure}[h]
\begin{subfigure}{0.3\textwidth}
\includegraphics[width=0.85\linewidth]{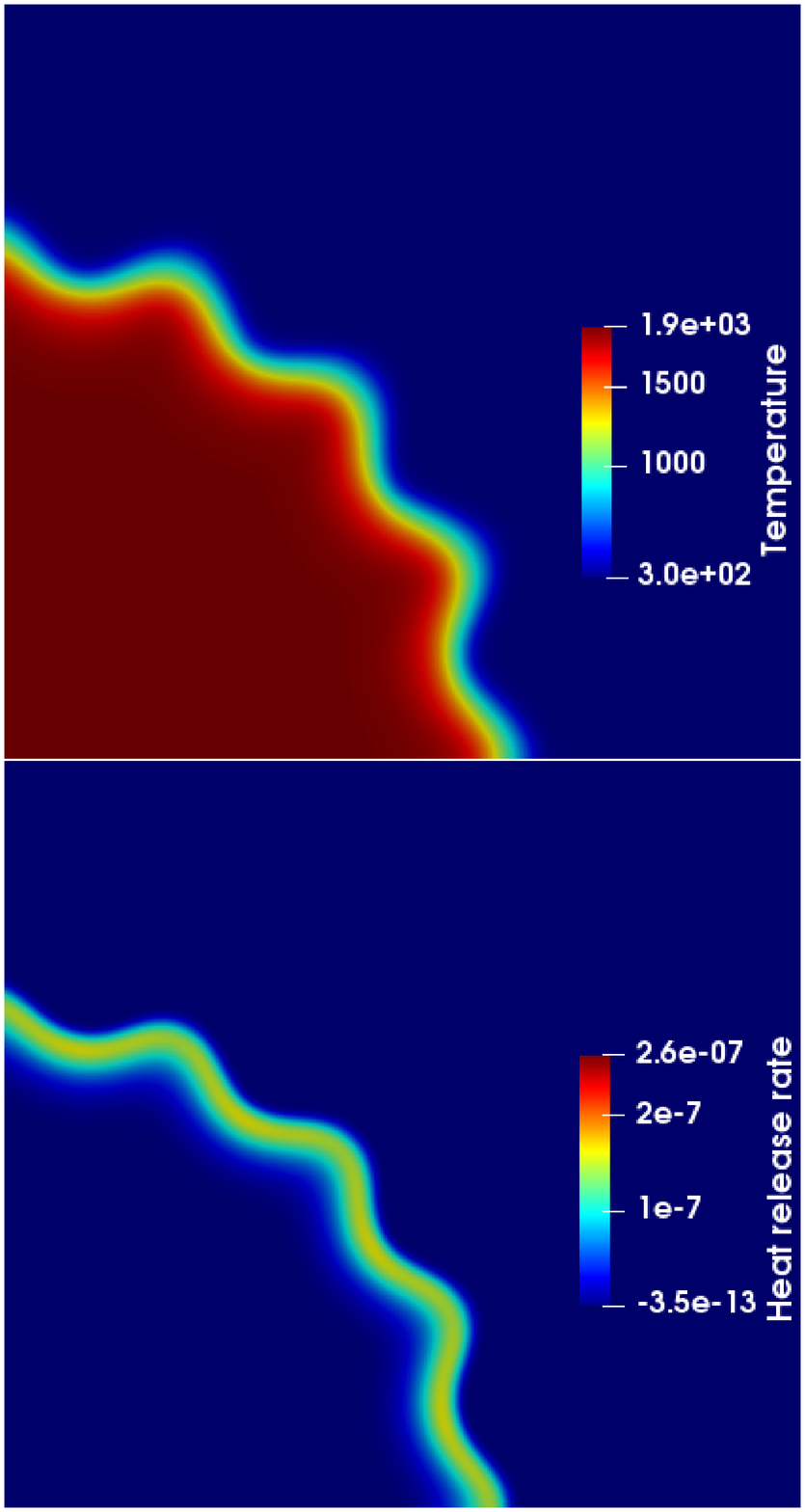} 
\caption{$t=0.024 \tau$.}
\label{fig:leftP}
\end{subfigure}
\begin{subfigure}{0.3\textwidth}
\includegraphics[width=0.85\linewidth]{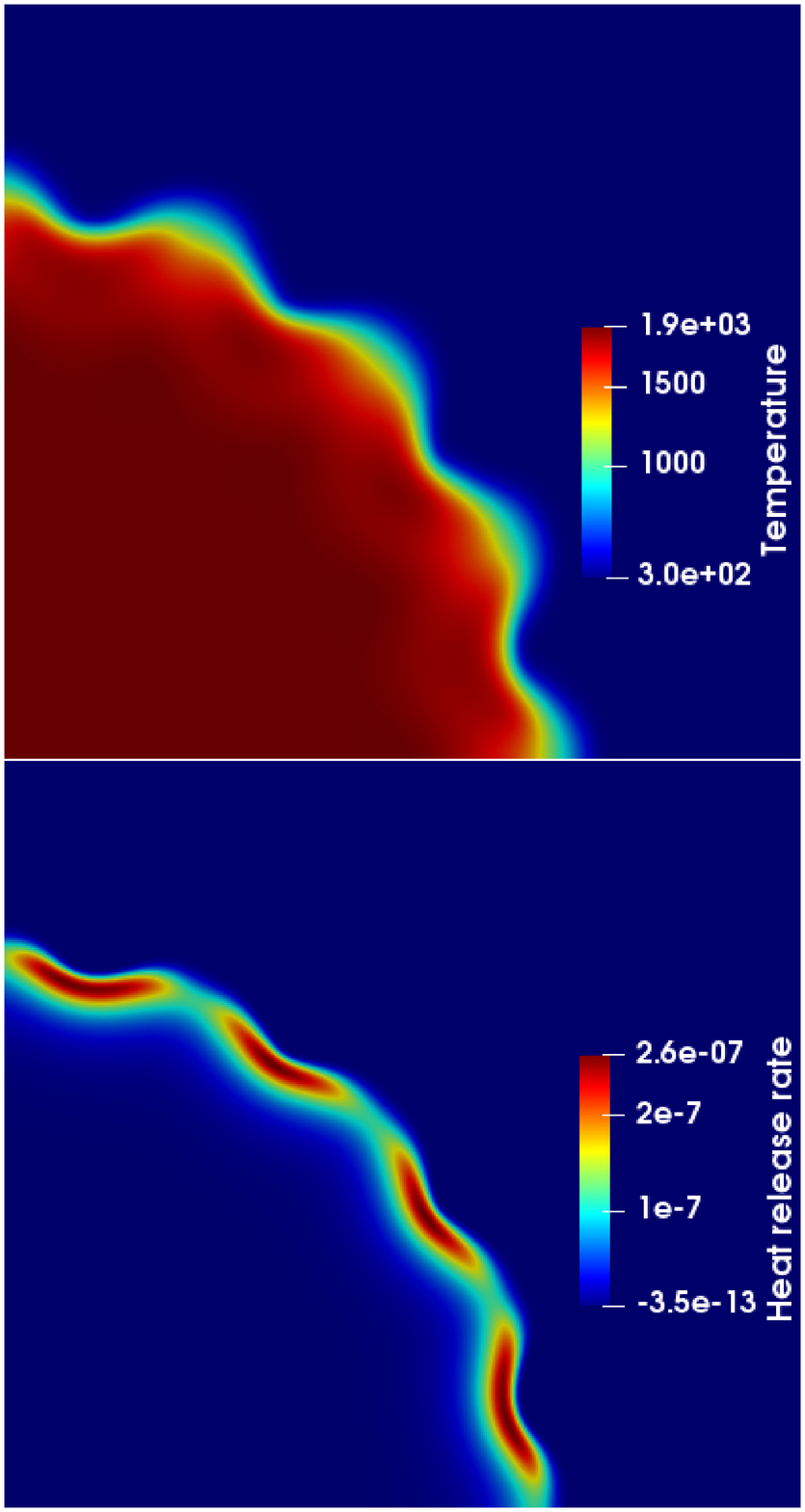}
\caption{$t=0.082 \tau$.}
\label{fig:rightP}
\end{subfigure}
\begin{subfigure}{0.4\textwidth}
\includegraphics[width=1.0\linewidth]{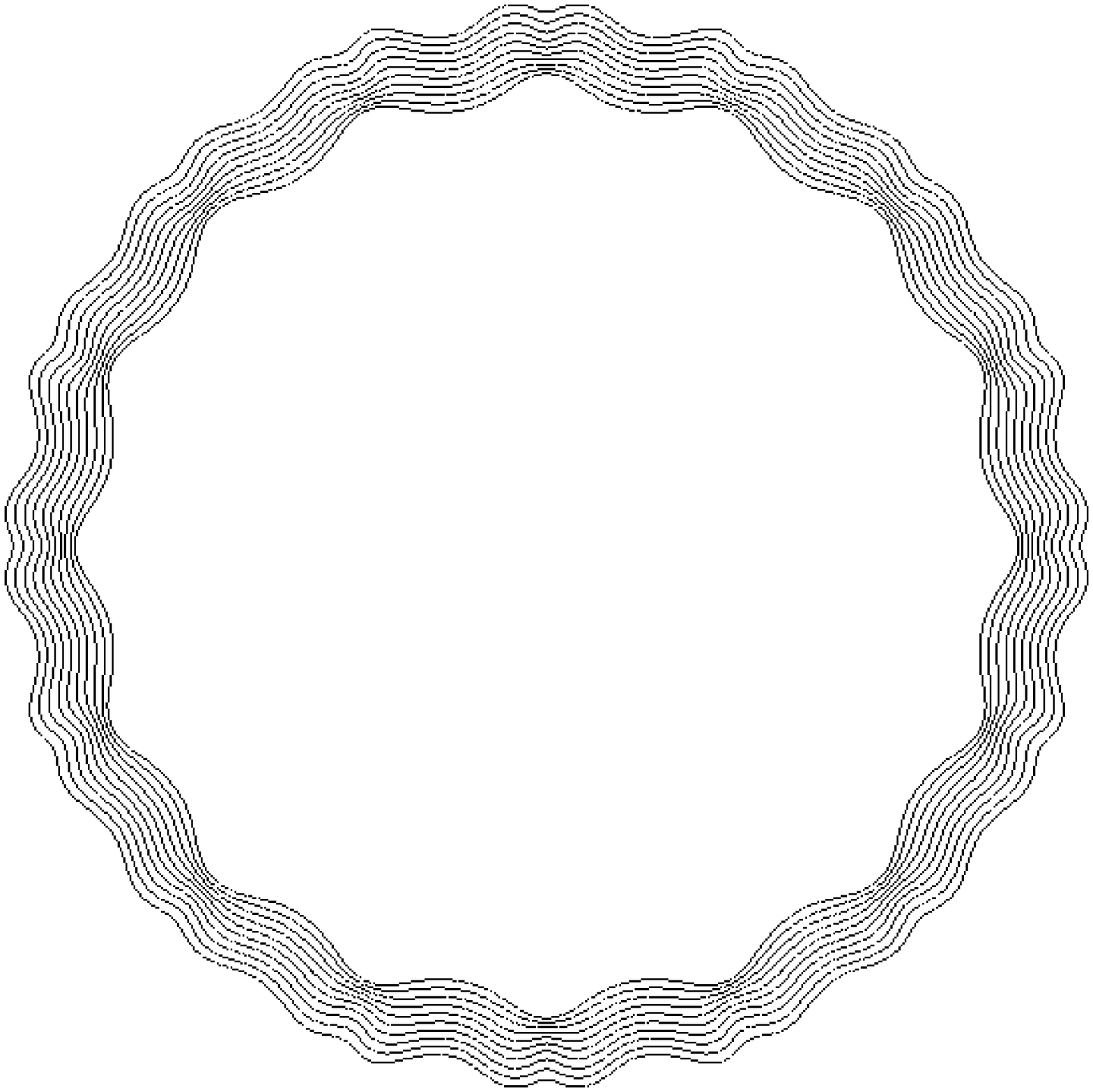}
\caption{Line contours of $T=1510.28 K$ form $t=0.041 \tau$ to $t=0.115 \tau$. The domain has been reflected about the left and the bottom edge for plotting.}
\label{fig:lineContours}
\end{subfigure}
\caption{Contours of temperature and heat release rate.}
\label{fig:imageP}
\end{figure}

Next, we study the response of this setup to a deterministic perturbation to validate the model with the Direct Numerical Simulation (DNS) of \cite{altantzis_direct_2015}. 
The initial circular profile of the flame is perturbed with a sinusoidal profile,
\begin{align}
    R(\theta) = R_{\rm ig} (1 + A_0 \cos ( 4 n_0 \theta ) ),
    \label{eq:cflamePerturbation}
\end{align}
where $n_0=4$ corresponds to the number of modes of the perturbation per $\pi/2$ sector of the flame and $A_0=0.05$ is the amplitude of the perturbation. The evolution of the perturbation is shown in Fig.\ \ref{fig:imageP}. {The heat release rate, $\dot h^c=-\sum_{a=1}^M H_a \dot \rho_a^c$, is a measure of the reactivity of the mixture.} As it is evident in Fig.\ \ref{fig:leftP}, during the initial stages of the evolution, the perturbed modes are continuous and the heat release rate is uniform along the circumference of the flame. As explained in \cite{altantzis_direct_2015}, the reactivity and therefore the heat release rate reduces at the crest due to diffusion and more consumption of the deficient reactant. This, along with the hydrodynamic instability due to the density ratio and the thermal-diffusive instability due to the heat and mass imbalance of the deficient reactant leads to splitting of the peak of the crests into smaller cells, as it is visible in Fig.\ \ref{fig:rightP}. A snapshot of the temperature contours over time shown in Fig.\ \ref{fig:lineContours} verifies that the splitting of the flame indeed occurs from crests. Therefore, the splitting stems from the deterministic perturbation as expected, and not because of numerical noise. The mean radius of the flame is calculated by integrating along the flame front circumference,
\begin{align}
\bar{R}={A}^{-1} \int R \; dA.
\label{eq:rmean}
\end{align}
Here $A$ is the circumferential length and $R$ is the distance of the mean temperature isoline from the centre. On fitting $\bar{R}=a t^{\alpha}$, the growth exponent was found to be $\alpha=1.16$, in agreement with the results from DNS in the literature wherein the value of the exponent was found to be between almost linear \cite{altantzis_direct_2015} and $1.25$ \cite{liberman_self-acceleration_2004}. 
The local displacement speed \cite{altantzis_numerical_2013,altantzis_direct_2015} is calculated as,
\begin{align}
S_d = \frac{1}{\rho C_p \lvert \nabla T \rvert } \left[  -\sum_{a=1}^M  H_a \dot \rho_a^c + \nabla \cdot (\lambda \nabla T) - \rho \left( \sum_{a=1}^M C_{a,p} Y_a \delta \bm{u}_a \right) \cdot \nabla T \right].
\label{eq:sd}
\end{align}
With the local flame normal $\bm{n}=-\nabla T/ \lvert \nabla T \rvert$, the absolute propagation speed is calculated as $S_a = S_d + \bm{u} \cdot \bm{n}$. The density weighted displacement speed is defined as $\hat{S_d}=\rho S_d/\rho_u$, where $\rho_u$ is the density of the unburnt mixture. The flame speeds are calculated as a mean over the flame interface isoline of $T=3 T_{\rm u}$ in a way similar to equation (\ref{eq:rmean}). After the initial transients, the absolute propagation speed was found to reach a value of $6.2 S_L$ whereas the density weighted displacement speed was found to fluctuate about $1.3 S_L$. The corresponding values from the DNS results \cite{altantzis_direct_2015} are $7 S_L$ and $1.5 S_L$ respectively. The difference could be attributed to a number of factors including the type of grid, resolution, type of diffusion model etc. Overall, the results agree well with the DNS \cite{altantzis_numerical_2013,altantzis_direct_2015}. 
 
\section{Conclusion}
\label{sec:conclusion}
In this paper, we proposed a lattice Boltzmann framework to simulate reactive mixtures. The novelty of the model lies in the fact that temperature and energy changes due to chemical reaction are handled naturally without the need of additional ad-hoc modelling of the heat of reaction. This was possible because of the thermodynamic consistency of the underlying multi-component model \cite{sawant_consistent_2021}, which was extended to compressible reactive mixtures. 
The species interaction is modelled through the Stefan--Maxwell diffusion mechanism which has been extended in this work to accommodate for the creation and destruction rates of the species due to chemical reaction. 
%Thus, the proposed framework is fully coupled through composition, momentum, pressure and enthalpy and is therefore free of any passive scalars.
Computational efficiency has been achieved through reduced description of energy which makes it possible to describe the physical system by only $M+2$ kinetic equations instead of $2M$ kinetic equations while retaining necessary physics such as the inter-diffusion energy flux. The model has been realized on the standard $D3Q27$ lattice, which not only reduces the computational costs compared to multispeed approaches but also possesses a wide temperature range, which is crucial for combustion applications.  

The proposed model was validated in one and two dimensions with the $9$-species $21$ steps detailed hydrogen-air reaction mechanism. The accuracy of the model was assessed by calculating the burning velocity of a premixed hydrogen-air mixture in $1D$. The calculated flame speed agrees well with the results in the literature. The ability of the model to capture complex physics was tested by simulating a $2D$ expanding circular flame. The circular flame simulation exhibited good isotropy and low numerical noise. The setup was then subjected to monochromatic perturbations in order to study the evolution of the perturbed flame.
Good agreement with DNS simulations demonstrates viability of the proposed LBM for complex reactive flows.

\enlargethispage{20pt}

\aucontribute{ N.S. implemented the model, ran the simulations and wrote the first draft of the manuscript. B.D. and I.V.K. supervised the project. 
All authors contributed to conceptualization of the model as well as writing, reading and approving the paper.
}

\competing{The authors declare that they have no competing interests.}

\funding{This work was supported by the European Research Council grant No. 834763-PonD.}

\ack{Computational resources at the Swiss National  Super  Computing  Center  CSCS  were  provided  under grant No. s897. 
Authors thank Ch. Frouzakis at ETHZ for discussions about the circular expanding flame. }

%%%%%%%%%% Insert bibliography here %%%%%%%%%%%%%%

%\bibliographystyle{vancouver}
%\bibliography{library}

\printbibliography

\end{document}